# On the Possibility of Generating Harmonics of the Electron Plasma Frequency in the Solar Atmosphere due to Explosive Instability in a System of Interpenetrating Electron and Ion Flows

V. V. Fomichev, S. M. Fainshtein, and G. P. Chernov

*Pushkov Institute of Terrestrial Magnetism, Ionosphere, and Radio Wave Propagation,
Russian Academy of Sciences, Moscow, 142191 Russia*
*e-mail: gchernov@izmiran.ru*

**Abstract**—An alternative mechanism is proposed for the generation of harmonics of the electron plasma frequency due to the development of explosive instability in a system of interpenetrating electron and proton flows in the solar atmosphere. The efficiency of the new mechanism in comparison with the earlier discussed mechanisms involving multistage processes of nonlinear interaction of waves in plasma is determined. It is shown that the development of explosive instability can lead to the excitation of the second and third harmonics of the plasma frequency with comparable amplitudes.

1. INTRODUCTION

Solar radio-frequency bursts in the meter wavelength band are, as a rule, very unsteady, short-term, and intense radiative phenomena with an effective radiation temperature ($T_{\text{eff}} \geq 10^{10}$ K) much higher than the temperature of the solar corona ($T \approx 10^6$ K). Presently, they are explained by a plasma mechanism according to which plasma waves are first excited at frequencies close to the electron plasma frequency in the solar corona and, then, these waves transform by one or another mechanism into electromagnetic waves, which can leave the generation region [1]. Typical examples of such bursts are type-II and type-III solar radio bursts. Type-III radio bursts, which are rapidly drifting in frequency (from high to low), are generated by fast electron flows propagating at an average velocity of $\sim c/3$ (where $c$ is the speed of light in vacuum) through the solar corona into the interplanetary space up to the Earth's orbit. Since the plasma density monotonically decreases with distance from the Sun, the fast electrons propagating in such a medium excite plasma waves at successively lower frequencies, which causes a negative drift of the radio frequency in such bursts.

Type-II radio bursts, which are slowly drifting in frequency, are associated with the propagation of shock waves in the solar corona and interplanetary space [2–5]. Under the conditions of the solar corona, where collisions of plasma particles are insignificant, the shock waves are collisionless. According to the theory of weak collisionless shock waves (with Mach numbers of $M < 2$), the front of a shock wave propagating in the solar corona perpendicular to the magnetic field has an oscillatory structure, i.e., comprises a sequence of compression solitons [6]. Due to the nonuniformity of the magnetic field in solitons, a relative drift of electrons and ions in the plane of the shock front arises, the velocity of which depends on the shock wave intensity. If the Mach number exceeds the critical value, $M_{cr} = 1 + \frac{3}{4}(8\pi n_0 \kappa T_e H_0^{-2})^{1/3}$, then the relative drift velocity of electrons and ions exceeds the electron thermal velocity. Here, $n_0$, $H_0$, and $T_e$ are the plasma density, magnetic field, and electron temperature ahead of the shock front, respectively. In this case, Buneman instability of plasma oscillations develops within the shock front, which leads to the generation of electromagnetic waves.

The dynamic spectra (time variation in the frequency spectra) of both types of radio bursts have a diverse fine structure. A typical feature of these spectra is their harmonic structure, i.e., the presence in them of simultaneously drifting radio emission bands at frequencies close to the electron plasma frequency and its second harmonic. Such a harmonic structure of radio bursts indicates that nonlinear processes in the solar atmosphere play an important role in the mechanism of radio burst generation. In particular, electromagnetic radiation at frequencies close to the plasma frequency was attributed to the scattering of plasma waves by density fluctuations (or to their interaction with ion-acoustic waves), while electromagnetic radiation at frequencies close to the second harmonic of the plasma frequency was explained by the interaction between plasma waves excited by a fast electron flow and oppositely directed plasma waves arising due to either stimulated scattering of excited plasma waves by ions or their decay instability. Higher order nonlinear processes were involved in [1, 5, 7] to explain the generation of type-III radio bursts at the third harmonic of the electron plasma frequency. In particular, in [7], generation of electromagnetic radiation at the third harmonic was explained in two ways: (i) as a result of the merging of three plasma waves (process $l_1 + l_2 + l_3 \rightarrow t$) or (ii) by the combination interaction of two plasma waves (process $l_1 + l_2 \rightarrow t_{II}$) and the subsequent interaction of the resulting electromagnetic radiation with a plasma wave (process $l_3 + t_{II} \rightarrow t_{III}$). Here, $l$ and $t$ refer to plasma and electromagnetic waves, respectively. Estimates of the efficiencies of these two processes evidence in favor of the twostep mechanism of third-harmonic generation.

These processes give a satisfactory explanation if the third harmonic generated due to the third stepwise process of merging is much weaker than the first two. However, according to observations, the third harmonic can be comparable in the intensity with the first two. In this case, the appearance of the third-harmonic radiation with an intensity comparable to the intensity of the second harmonic requires the presence of intense plasma waves, i.e., plasma turbulence should be strong. It is worth noting that, in some cases, the dynamic spectra of type-II radio bursts indicate the presence of radiation at harmonics with numbers $n = 3$ and 4 (and, possibly, 5) [7–9], the intensities of the higher harmonics being nearly the same. Since the efficiencies of the above mechanisms sharply decrease when passing to the higher harmonics of the plasma frequency, the mechanism of generation of higher harmonics in solar radio emission remains unclear. One possible solution to this problem can be the development of explosive instability in the nonlinear interaction of three types of waves in the radiation source. In [10], a mechanism was considered according to which harmonics of the electron plasma frequency were generated as a result of the development of explosive instability in plasma penetrated by an electron flow. It was shown that this mechanism makes it possible to explain the emergence of higher harmonics with comparable intensities in the solar radio emission under the conditions existing in the sources of type-III radio burst.

In this paper, we consider the possibility of excitation of electromagnetic radiation at harmonics of the electron plasma frequency due to the development of explosive instability in a system of interpenetrating electron and ion flows. As noted above, such a situation can occur at the front of a collisionless shock wave due to the gradient drift of electrons and ions, while the plasma waves excited due to instability can be a source of type-II radio bursts. Conditions for the development and saturation of explosive instability are revealed. The amplitudes of harmonics of the electron plasma frequency for the solar plasma conditions are estimated qualitatively. Such a mechanism for the generation of radio emission at the third and higher harmonics of the electron plasma frequency in type-II solar radio bursts has not been considered earlier.

2. BASIC EQUATIONS AND SYNCHRONISM CONDITIONS

As was pointed out in the previous section, due to the regular nonuniformity of the magnetic field at the front of a weak ($M < 2$) collisionless shock wave propagating in the solar corona perpendicular to the magnetic field, a relative drift of electrons and ions in the plane of the front takes place. Under the condition $M > M_{cr}$, the velocity of the relative drift exceeds the electron thermal velocity, which leads to the development of Buneman instability and excitation of plasma oscillations inside the shock front. The system of one-dimensional ($\partial/\partial x \neq 0$) quasihydrodynamic equations describing the motion of weakly relativistic electrons and positive ions (protons) in the self-consistent electric field has the form [11]:

$$\frac{\partial E}{\partial x} = 4\pi e\,(\rho_e - \rho_i);\; \frac{\partial \rho_i}{\partial t} + V_{0i}\frac{\partial \rho_i}{\partial x} + N_i\frac{\partial v_i}{\partial x} = -\mu\frac{\partial}{\partial x}(\rho_i v_i);$$

$$\frac{\partial \rho_e}{\partial t} + V_{0e}\frac{\partial \rho_e}{\partial x} + N_e\frac{\partial v_e}{\partial x} = \mu\frac{\partial}{\partial x}(\rho_e v_e);$$

$$\frac{\partial v_i}{\partial t} - V_{0i}\frac{\partial v_i}{\partial x} + \frac{e}{m_i}E = -\mu\, v_i\frac{\partial v_i}{\partial x} \tag{1}$$

$$\frac{\partial v_e}{\partial t} + V_{0e}\frac{\partial v_e}{\partial x} - \frac{e}{m_e}E = -\mu\, v_e\frac{\partial v_e}{\partial x} + \frac{\mu\, eE}{m_e c^2}v_e^2.$$

Here, $x$ is the coordinate in the plane of the shock front, along which the current flows; $E$ is the electric field; $m_e$ and $m_i$ are the masses of an electron and ion, respectively; $V_{0i}$ and $V_{0e}$ are the equilibrium velocities of ions and electrons; $N_e$ and $N_i$ are the electron and ion equilibrium densities; $v_e$, $v_i$, $\rho_e$, and $\rho_i$ are the deviations of the electron and ion velocities and densities from their equilibrium values; $c$ is the speed of light in vacuum; $e$ is the electron charge; and the parameter $\mu \sim |\rho_e/N_e| \sim |\rho_i/N_i| \sim |v_e/V_{0e}| \sim |v_i/V_i|$ ($\mu \ll 1$) is introduced to denote the smallness of the right-hand sides of Eqs. (1). System of equations (1) contains quadratic and cubic nonlinearities. Note that, in the equation for $v_e$, we take into account weak electron relativism, which, as will be shown below, stabilizes the explosion.

Let us consider wave processes in which the variable quantities depend on the coordinate and time as $\sim\exp(i\omega t - ikx)$, where $\omega$ and $k$ are the circular frequency and wavenumber of the wave perturbation, respectively. After substituting perturbations in such a form into system of equations (1), we obtain the following dispersion relation describing the normal modes of the plasma system under study:

$$1 - \omega_{0i}^2\,(\omega + kV_{0i})^{-2} - \omega_0^2(\omega - kV_{0e})^{-2} = 0, \tag{2}$$

where $\omega_{0i}^2 = 4\pi e^2 N_i/m_i$ and $\omega_0^2 = 4\pi e^2 N_e/m_e$ are the ion and electron plasma frequencies, respectively. Let us consider the resonance three-wave interaction of normal waves in this plasma system. In this interaction, the following laws of conservation of energy and momentum should be satisfied [12]:

$$\omega_3 = \omega_1 + \omega_2, \quad k_3 = k_1 + k_2. \tag{3}$$

Using dispersion relation (2) and conditions (3), we obtain the parameters of the resonant triplet

$$\omega_3 \approx \omega_2 + \omega_{0i}, \quad \omega_2 = n\omega_0 - \omega_{0i},$$
$$k_3 \sim n\omega_0/V_{0e}, \quad k_2 \sim k_3, \quad \omega_1 \sim \omega_{0i}, \tag{4}$$
$$k_1 \sim \omega_{0i}/V_{0i},$$

where $\omega_1$ is the Buneman mode and $\omega_2$ and $\omega_3$ are the beam modes (with $n$ being the mode number). Having determined the energies of the interacting modes, we find that a wave with the frequency $\omega_3$ (the third plasma frequency harmonic) has a negative energy, while the two others ($\omega_1$ and $\omega_2$) have positive energies; i.e., the system admits an explosive solution, which is stabilized by the cubic nonlinearity in the last of Eqs. (1).

3. ANALYSIS OF EQUATIONS FOR THE WAVE AMPLITUDES

We represent all variables in Eqs. (1) in the form $\sim a_j \exp(i\omega_j t - ik_j x)$ and expand the nonlinear terms in Eqs. (1) in a Taylor series up to cubic terms (the cubic terms are related to the dependence of the mass of a moving electron on the velocity $V_{0e}$). Then, using the standard technique [13, 14], we obtain reduced equations for the slowly varying complex wave amplitudes $a_j(\mu t, \mu x)$, ($E_{xj} = a_j(\mu t, \mu x)\exp(i\omega_j t - ik_j x)$, $j = 1, 2, 3$). Here, we consider the one-dimensional resonance interaction of waves ($\partial/\partial x \neq 0$, $\partial/\partial y = 0$, $\partial/\partial z = 0$). Following the procedure used in [15], we obtain

$$\frac{\partial a_3}{\partial t} + V_{\text{gr}3}\frac{\partial a_3}{\partial x} = \sigma_3 a_1 a_2 + i a_3 \alpha_3 \sum_{j=1}^{3}|a_j|^2,$$

$$\frac{\partial a_{2,1}}{\partial t} + V_{\text{gr}2,1}\frac{\partial a_{2,1}}{\partial x} = \sigma_{2,1} a_3 a_{1,2}^*,$$
(5)

where

$$\sigma_1 \approx \frac{e}{mV_{0e}\beta^4}, \quad \sigma_{2,3} \approx \frac{2en}{mV_{0e}}, \quad \alpha_3 \approx \frac{e}{4m^2c^2\omega_0},$$

$$\beta \sim \frac{V_{0e}}{V_{0i}}, \quad V_{0e} \approx \frac{c}{3},$$
(6)

and $V_{\text{gr}1,2,3}$ are the group velocities of the modes with n = 1, 2, and 3.

According to formulas (6), the signs of the coefficients $\sigma_{1,2,3}$ are the same; therefore, the wave amplitudes can increase simultaneously. This means that, in the system under consideration, explosive instability can develop. For the convenience of solving Eq. (5), we represent the amplitudes $a_j$ in the form $a_j = u_j(\sigma_m\sigma_n)^{1/2}$ ($j = 1, 2, 3$). As a result, for waves with the same initial conditions $u_3(0) = u_2(0) = u_1(0)$, we obtain a solution to system of equations (5) for wave amplitudes in the spatially homogeneous regime ($\partial/\partial x = 0$, $\partial/\partial t \neq 0$). The maximum amplitude of the mode with the frequency $\sim n\omega_0$ ($n = 1, 2, 3$) can easily be obtained from Eq. (5),

$$a_n^{\max} \approx \frac{\sqrt{\sigma_1\sigma_2}}{\alpha_n},$$
(7)

where $\alpha$ is the coefficient in the term with cubic nonlinearity in Eq. (5). It is obtained by averaging the cubic term in the fourth of Eqs. (1). Since $\alpha_3 \gg \alpha_{1,2}$, the expression only for $\alpha_3$ is given in formulas (6).

The characteristic explosion time is $t^* = 1/u_0$, where $u_0$ is the initial normalized wave amplitude, which is related to $a_j(0)$ as $u_j = a_j/(\sigma_m \sigma_n)^{1/2}$. Hence, we have $t^*(\omega_3) = \sigma_2 \sigma_1/a_3(0)$, where $a_3(0)$ is the initial noise amplitude.

## 4. QUALITATIVE ESTIMATES

Let us estimate the amplitudes of a soliton at the frequency $n\omega_0$ ($n = 1, 2, 3$) for solar plasma with $N_e \sim 10^8$ cm$^{-3}$ ($\omega_0 \sim 10^9$ s$^{-1}$) at $V_0 \sim c/3$ and a constant magnetic field of $H_0 \sim 1$ G, which corresponds to an electron gyrofrequency of $\omega_H \sim 2 \times 10^7$ s$^{-1}$ (i.e., $\omega_0 \gg \omega_H$). It is easy to find from expression (7) that $a_3^{max} \approx \sqrt{n}\, 4mc^2\omega_0/V_0\beta^2 e$, i.e., $a_3^{max} \sim 70$ V/cm, $a_2^{max} \sim 56$ V/cm, and $a_1^{max} \sim 40$ V/cm. The closeness of the values of $a_{1,2,3}$ in spite of the relationship $\alpha_3 \gg \alpha_{1,2}$ is due to the fact that the inequality $\sigma_1 \gg \sigma_{2,3}$ is simultaneously satisfied (see formulas (6)). Taking the noise value $a_N \sim 10^{-3} a_3^{max}$ as the initial value of $a_3(0)$, we find that the development time of explosion is $t^* = 1/u_0 \sim 20$ μs.

Comparing the amplitudes of the plasma frequency harmonics ($n = 1, 2, 3$) obtained in [9] as a result of the development of explosive instability in plasma penetrated by a proton flow with the amplitudes of the harmonics generated by interpenetrating ion and electron flows, we find that the latter are higher. Thus, in the present paper, an alternative mechanism for the generation of harmonics of the electron plasma frequency by interpenetrating ion and electron flows has been revealed.

## 5. CONCLUSIONS

The above-discussed situation with the generation of electromagnetic radiation in a system of interpenetrating electron and proton flows is directly related to the generation of type-II solar radio bursts. At present, according to the most developed and recognized model, type-II radio bursts are generated by collisionless shock waves excited in the solar corona either at the fronts of coronal mass ejections or after solar bursts. In any case, at the front of a collisionless shock wave propagating perpendicular to the external magnetic field, Buneman instability develops due to the gradient drift of electrons and protons, which can lead to the generation of radio waves at the first and second harmonics of the electron plasma frequency [2–5].

It is also not excluded that some type-II bursts are associated with shock waves propagating along the magnetic field in the solar corona. In this case, modified Buneman instability can develop, which leads to the generation of plasma waves at the lower hybrid frequency.

These waves can heat the plasma at the shock wave front. At sufficiently large Mach numbers, fast electrons injected ahead of the front can excite coherent plasma waves, the scattering of which leads to type-II radio bursts.

The presence of bands at the third and higher plasma frequency harmonics in the dynamic spectra of type-II radio bursts requires an additional analysis of mechanisms for the generation of solar radio bursts. In [10], a mechanism for the generation of harmonics of the electron plasma frequency due to the development of explosive instability in plasma penetrated by an electron flow was considered; it was noted hat this mechanism might be responsible for the specific features of type-III radio bursts. In the present paper, we have analyzed the mechanism for the excitation of plasma frequency harmonics in type-II bursts. The mechanism based on explosive instability in a system of interpenetrating electron and proton flows, which actually exist at the front of a collisionless shock wave, turns out to be more efficient than the alternative mechanisms of harmonic generation in equilibrium plasma or mechanisms of three-wave interaction (see [7]). The reason for this is that, in the mechanism proposed here, modes with different $n$ are excited simultaneously. In addition, the amplitudes of the electric field harmonics provided by this mechanism (i.e., upon the interaction of electron flows with plasma) exceed their typical values in sources of type-III radio bursts by more than one order of magnitude due to the contribution of ion flows to the excitation of plasma turbulence. Conditions for the development of explosive instability and its saturation due to the nonlinear frequency shift caused by the dependence of the electron mass on velocity have been determined. Qualitative estimates show that the amplitudes of different plasma frequency harmonics generated in the solar plasma are comparable. The results of this study can be used to interpret type-II radio bursts generated in the solar corona. In addition, intense electromagnetic fluxes generated in the course of explosive instability can be used to interpret strong electromagnetic radiation emitted from the atmospheres of stars and pulsars [16].

To conclude, it should be noted that, here, we did not take into account the linear instability that can arise during Buneman instability. As was shown in [17], the linear instability is suppressed by strong RF fields; i.e., in our case, the powerful RF field arising in the course of explosive instability completely suppresses the linear instability.

## ACKNOWLEDGMENTS

This work was supported by the Russian Foundation for Basic Research, project no. 17-02-00308.